\documentclass[journal,draftclsnofoot,onecolumn]{IEEEtran}
\usepackage{graphicx}
\usepackage{amssymb}

\usepackage[caption=false,font=normalsize,labelfont=sf,textfont=sf]{subfig}
\usepackage{caption}
\renewcommand{\theequation}{\arabic{section}.\arabic{equation}}
\newtheorem{theorem}{Theorem}

\newtheorem{lemma}{Lemma}

\newtheorem{remark}{Remark}

\begin{document}

\title{Multiple Access Wiretap Channel with Noiseless Feedback}

\author{Bin~Dai
        and~Zheng~Ma
\thanks{B. Dai is with the
School of Information Science and Technology,
Southwest JiaoTong University, Chengdu 610031, China, and with
the State Key Laboratory of Integrated Services Networks, Xidian University, Xi$'$an, Shaanxi 710071, China,
e-mail: daibin@home.swjtu.edu.cn.}
\thanks{
Z. Ma is with the
School of Information Science and Technology,
Southwest JiaoTong University, Chengdu 610031, China,
e-mail: zma@home.swjtu.edu.cn.}
}

\maketitle

\begin{abstract}

The physical layer security in the up-link of the wireless communication systems is often modeled as the multiple access wiretap channel (MAC-WT), and
recently it has received a lot attention. In this paper, the MAC-WT has been re-visited by considering the situation that the legitimate receiver feeds
his received channel output back to the transmitters via two noiseless channels, respectively. This model is called the MAC-WT with
noiseless feedback. Inner and outer bounds on the secrecy capacity region of this feedback model are provided. To be specific,
we first present a decode-and-forward (DF) inner bound on the secrecy capacity region of this feedback model, and this bound
is constructed by allowing each transmitter to decode the other one's transmitted message from the feedback, and then each transmitter
uses the decoded message to re-encode his own messages, i.e., this DF inner bound allows
the independent transmitters to co-operate with each other.
Then, we provide a hybrid inner bound
which is strictly larger than the DF inner bound, and it is constructed by using the feedback as a tool not only to
allow the independent transmitters to co-operate with each other, but also to generate two secret keys respectively
shared between the legitimate receiver and the two transmitters. Finally, we give a sato-type outer bound on the secrecy capacity region of this feedback model.
The results of this paper are further explained via a
Gaussian example.

\end{abstract}

\begin{IEEEkeywords}
Multiple-access wiretap channel, noiseless feedback, secrecy capacity region.
\end{IEEEkeywords}

\section{Introduction \label{secI}}
\setcounter{equation}{0}

The physical layer security (PLS) was first investigated by Wyner in his landmark paper on the degraded wiretap channel \cite{Wy}.
Wyner's degraded wiretap channel model consists of one transmitter and two receivers (a legitimate receiver and an eavesdropper).
The transmitter sends a private message to the legitimate receiver via a discrete memoryless main channel,
and an eavesdropper eavesdrops the output of the main channel via another discrete memoryless wiretap channel.
We say that the perfect secrecy is achieved if no information about the private message is leaked to the eavesdropper.
The secrecy capacity $C_{s}$, which is the maximum reliable transmission rate with perfect secrecy constraint,
was characterized by Wyner \cite{Wy}, and it is given by
\begin{eqnarray}\label{c1}
&&C_{s}=\max_{p(x)}(I(X;Y)-I(X;Z)),
\end{eqnarray}
where $X$, $Y$ and $Z$ are the input of the main channel, output of the main channel and output of the wiretap channel, respectively,
and they satisfy the Markov chain
$X\rightarrow Y\rightarrow Z$. Here note that (\ref{c1}) holds under the degradedness assumption $X\rightarrow Y\rightarrow Z$, and the secrecy
capacity of the general wiretap channel (the wiretap channel without the degradedness assumption) was determined by
Csisz$\acute{a}$r and K\"{o}rner \cite{CK}.
The work of \cite{Wy} and \cite{CK}
lays a foundation for the PLS of
the practical communication systems.

Since Wozencraft et al. \cite{supp1} showed that the time-variant noisy two-way channels
can be used to provide noiseless feedback, whether this noiseless feedback helps to enhance the capacities of various communication channels
motivates the researchers to study the channels with noiseless feedback. Shannon first proved that the noiseless feedback
does not increase the capacity of a point-to-point discrete memoryless channel (DMC) \cite{coverx}. After that,
Cover et al. \cite{coverz, CG1} and Bross et al. \cite{lapidoth} showed that the capacity regions of several multi-user channels, such as multiple-access channel (MAC) and relay channel, can
be enhanced by feeding back the receiver's channel output to the transmitter over a noiseless channel.
Then, it is natural to ask: does the noiseless feedback from the legitimate receiver to the transmitter also help to enhance the secrecy capacity of
the wiretap channel? Ahlswede and Cai \cite{AC} answered this question by considering the wiretap channel with noiseless feedback.
Since the noiseless feedback is known by the legitimate receiver and the transmitter, and it is not available for the eavesdropper,
Ahlswede and Cai pointed out that the noiseless feedback can be used to generate a secret key shared only
between the transmitter and the legitimate receiver, and we can use this key to encrypt the transmitted messages.
Combining the idea of generating a secret key from the noiseless feedback with Wyner's random binning technique used in the achievability proof of (\ref{c1}),
Ahlswede and Cai showed that
the secrecy capacity $C_{sf}$ of the degraded wiretap channel with noiseless feedback is given by
\begin{eqnarray}\label{c2}
&&C_{sf}=\max_{p(x)}\min\{I(X;Y),I(X;Y)-I(X;Z)+H(Y|X,Z)\},
\end{eqnarray}
where $X$, $Y$ and $Z$ are defined the same as those in (\ref{c1}), and $X\rightarrow Y\rightarrow Z$
forms a Markov chain.
Comparing (\ref{c2}) with (\ref{c1}),
it is easy to see that the noiseless feedback increases the secrecy capacity of the degraded
wiretap channel. Other related works on the wiretap channel with noiseless feedback are in \cite{AFJK}-\cite{dai2}.

\begin{figure}[htb]
\centering
\includegraphics[scale=0.5]{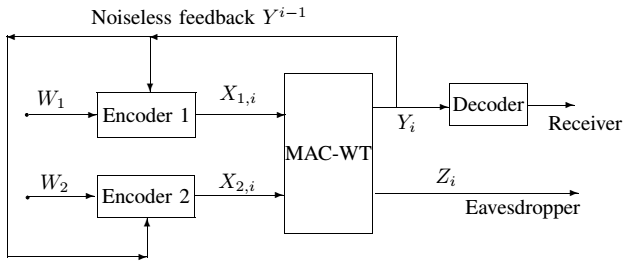}
\caption{The multiple-access wiretap channel with noiseless feedback}
\label{f1}
\end{figure}

In recent years, the PLS in the up-link of wireless communication system receives a lot attention, see
\cite{TY1}-\cite{WB}. These work extends Wyner's wiretap channel
to a multiple access situation: the multiple-access wiretap channel (MAC-WT). Bounds on the secrecy capacity region of MAC-WT
are provided in \cite{TY1}-\cite{WB}.
In order to investigate whether the noiseless feedback from the legitimate receiver to the transmitters helps to enhance
the secrecy capacity region of the MAC-WT, in this paper, we study the MAC-WT with
noiseless feedback, see Figure \ref{f1}. We first present a DF inner bound on the secrecy capacity region of the model of Figure \ref{f1},
and this bound is constructed by using the DF strategy of the MAC-WT with noisy feedback \cite{tang}, where each transmitter of the MAC decodes
the other one's transmitted message from
the noisy feedback and then uses it to re-encode his own messages. Second, note that the noiseless feedback can not only be used to re-encode the
messages of the transmitters, but also be used to generate secret keys to encrypt the transmitted messages, thus
we present a hybrid inner bound on the secrecy capacity region of the model of Figure \ref{f1} by combining Ahlswede and Cai's idea of generating
a secret key from the noiseless feedback \cite{AC} with the DF strategy used in \cite{tang}, and 
we show that this hybrid inner bound is strictly larger than the DF inner bound. Third, we present a sato-type outer bound 
on the secrecy capacity region of the model of Figure \ref{f1}. Finally,
the results of this paper are further explained via a Gaussian
example.

The rest of this paper is organized as follows. In
Section \ref{secII}, we show the definitions, notations and the main results of the model of Figure \ref{f1}.
An Gaussian example of the model of Figure \ref{f1} is provided in Section \ref{secIII}.
Final conclusions are presented in Section \ref{secIV}.

\section{Model description and the main result}\label{secII}
\setcounter{equation}{0}

\emph{Basic notations:} We use the notation $p_{V}(v)$ to denote
the probability mass function $Pr\{V=v\}$, where $V$ (capital letter) denotes the random variable, $v$ (lower case letter)
denotes the real value of the random variable $V$. Denote the alphabet in which the random variable $V$ takes values by $\mathcal{V}$ (calligraphic letter).
Similarly, let $U^{N}$ be a random vector $(U_{1},...,U_{N})$, and $u^{N}$ be a vector value
$(u_{1},...,u_{N})$. In the rest of this paper, the log function is taken to the base 2.

\emph{Definitions of the model of Figure \ref{f1}:}

Let $W_{1}$, uniformly distributed over the finite alphabet $\mathcal{W}_{1}=\{1,2,...,M_{1}\}$, be the message sent by the transmitter 1.
Similarly, let $W_{2}$, uniformly distributed over the finite alphabet $\mathcal{W}_{2}=\{1,2,...,M_{2}\}$, be the message sent by the transmitter 2.

The inputs of the channel are $x_{1}^{N}$ and $x_{2}^{N}$, while the outputs are $y^{N}$ and $z^{N}$. The channel is discrete memoryless, i.e.,
at the $i$-th time, the channel outputs $Y_{i}$ and $Z_{i}$ depend only on $X_{1,i}$ and $X_{2,i}$, and thus we have
\begin{eqnarray}\label{e201}
&&P_{Y^{N},Z^{N}|X_{1}^{N},X_{2}^{N}}(y^{N},z^{N}|x_{1}^{N},x_{2}^{N})\nonumber\\
&&=\prod_{i=1}^{N}P_{Y,Z|X_{1},X_{2}}(y_{i},z_{i}|x_{1,i},x_{2,i}).
\end{eqnarray}

Since $y^{N}$ can be fed back to the transmitters via a noiseless feedback channel, at the $i$-th time, the channel input $X_{j,i}$ ($j=1,2$) is given by
\begin{equation}\label{e202}
X_{j,i}=
\left\{
\begin{array}{ll}
f_{j,i}(W_{j}), & i=1\\
f_{j,i}(W_{j},Y^{i-1}), & 2\leq i\leq N.
\end{array}
\right.
\end{equation}
Here note that the $i$-th time channel encoder $f_{j,i}$ ($j=1,2$) is a stochastic encoder, and
the transmission rates of the messages $W_{1}$ and $W_{2}$ are $\frac{\log M_{1}}{N}$ and $\frac{\log M_{2}}{N}$, respectively.

The decoder is a mapping
$
\psi: \,\, \mathcal{Y}^{N}\rightarrow \mathcal{W}_{1}\times \mathcal{W}_{2},
$
with input $Y^{N}$ and outputs $\hat{W}_{1}$, $\hat{W}_{2}$.
The average probability of error $P_{e}$ is denoted by
\begin{equation}\label{e204}
P_{e}=\frac{1}{M_{1}M_{2}}\sum_{i=1}^{M_{1}}\sum_{j=1}^{M_{2}}Pr\{\psi(y^{N})\neq (i,j)|(i,j)\,\,\mbox{sent}\}.
\end{equation}
The eavesdropper's equivocation to the messages $W_{1}$ and $W_{2}$ is defined as
\begin{equation}\label{e205}
\Delta=\frac{1}{N}H(W_{1},W_{2}|Z^{N}).
\end{equation}

A positive rate pair $(R_{1}, R_{2})$ is called achievable
with weak secrecy if, for any small positive $\epsilon$, there exists an
$(M_{1}, M_{2}, N, P_{e})$ code such that
\begin{eqnarray}\label{e205}
&&\frac{\log M_{1}}{N}\geq R_{1}-\epsilon,
\frac{\log M_{2}}{N}\geq R_{2}-\epsilon, \Delta\geq R_{1}+R_{2}-\epsilon, \,\,\,\,P_{e}\leq \epsilon.
\end{eqnarray}
Here we note that $\Delta\geq R_{1}+R_{2}-\epsilon$ also ensures
$\frac{1}{N}H(W_{t}|Z^{N})\geq R_{t}-\epsilon$ for $t=1, 2$, and the proof is in \cite[p. 609]{tang}.
The secrecy capacity region $\mathcal{C}_{s}$ of the model of Figure \ref{f1} is a set composed of all rate pairs $(R_{1}, R_{2})$ satisfying (\ref{e205}).
The following Theorem \ref{T2} and Theorem \ref{T1} show two inner bounds on $\mathcal{C}_{s}$, and Theorem \ref{Tk} shows an outer bound
on $\mathcal{C}_{s}$.

\begin{theorem}\label{T2}
For the discrete memoryless MAC-WT with noiseless feedback,
an inner bound $\mathcal{C}^{DF}_{s}$ on the secrecy capacity region $\mathcal{C}_{s}$ is given by
\begin{eqnarray*}
&&\mathcal{C}^{DF}_{s}=\{(R_{1}\geq 0,R_{2}\geq 0): R_{1}\leq I(X_{1};Y|X_{2},U)\\
&&R_{2}\leq I(X_{2};Y|X_{1},U)\\
&&R_{1}+R_{2}\leq \min\{I(X_{1},X_{2};Y), I(X_{1};Y|X_{2},U)+I(X_{2};Y|X_{1},U)\}-I(X_{1},X_{2};Z)\},
\end{eqnarray*}
for some distribution
\begin{eqnarray}\label{the1}
&&P_{Z,Y|X_{1},X_{2}}(z,y|x_{1},x_{2})\cdot P_{X_{1}|U}(x_{1}|u)\cdot P_{X_{2}|U}(x_{2}|u)\cdot P_{U}(u).
\end{eqnarray}
\end{theorem}
\begin{IEEEproof}

In the MAC-WT with noisy feedback \cite{tang}, the legitimate receiver's channel output $Y$ is sent to the transmitters via two noisy feedback channels,
and the outputs of the noisy feedback channel are $Y_{1}$ and $Y_{2}$.
Substituting $Y_{1}=Y_{2}=Y$ (which implies the feedback channel is noiseless)
into \cite[Theorem 2]{tang}, the DF inner bound $\mathcal{C}^{DF}_{s}$ for the model of Figure \ref{f1} is obtained, and the proof of
$\mathcal{C}^{DF}_{s}$ is along the lines of that of \cite[Theorem 2]{tang} (the full DF inner bound on the secrecy capacity region of the MAC-WT with noisy
feedback), and thus we omit the proof here.
\end{IEEEproof}

\begin{remark}\label{R1.1}
In \cite[Theorem 1]{tang}, Tang et al. also provide a partial DF inner bound on the secrecy capacity region of the MAC-WT with noisy feedback.
Substituting $Y_{1}=Y_{2}=Y$ into \cite[Theorem 1]{tang}, and using Fourier-Motzkin elimination (see, e.g., \cite{lall}) to eliminate $R_{10}$,
$R_{12}$, $R_{20}$ and $R_{21}$, it is not difficult to show that the partial DF inner bound $\mathcal{C}^{PDF}_{s}$ of the model of Figure \ref{f1}
is exactly the same as
the DF inner bound $\mathcal{C}^{DF}_{s}$ shown in Theorem \ref{T2}.
\end{remark}

\begin{theorem}\label{T1}
For the discrete memoryless MAC-WT with noiseless feedback,
an inner bound $\mathcal{C}^{in}_{s}$ on the secrecy capacity region $\mathcal{C}_{s}$ is given by
\begin{eqnarray*}
&&\mathcal{C}^{in}_{s}=\{(R_{1}\geq 0,R_{2}\geq 0): R_{1}\leq I(X_{1};Y|X_{2},U)\\
&&R_{2}\leq I(X_{2};Y|X_{1},U)\\
&&R_{1}+R_{2}\leq \min\{I(X_{1},X_{2};Y), I(X_{1};Y|X_{2},U)\\
&&+I(X_{2};Y|X_{1},U)\}-I(X_{1},X_{2};Z)\\
&&+\min\{I(X_{1},X_{2};Z),H(Y|Z,X_{1},X_{2})\}\},
\end{eqnarray*}
for some distribution satisfying (\ref{the1}).
\end{theorem}
\begin{IEEEproof}

The hybrid inner bound $\mathcal{C}^{in}_{s}$ is constructed by combining Ahlswede and Cai's idea of generating
a secret key from the noiseless feedback \cite{AC} with the DF strategy used in \cite[Theorem 2]{tang},
and it is
achieved by the following key steps:
\begin{itemize}
\item For the transmitter 1, split the transmitted message $W_{1}$ into $W_{1,0}$ and $W_{1,1}$, and let
$W_{1}^{*}$ be a dummy message randomly generated by the transmitter 1, and it is used to confuse the eavesdropper.
Analogously, for the transmitter 2, split the transmitted message $W_{2}$ into $W_{2,0}$ and $W_{2,1}$,
and let $W_{2}^{*}$ be a dummy message randomly generated by the transmitter 2, and it is used to confuse the eavesdropper.

\item The messages $W_{1}$ and $W_{2}$ are transmitted through $n$ blocks, and in block $i$ ($2\leq i\leq n$), when each transmitter receives
the noiseless feedback, he tries to decode the other transmitter's message (including the transmitted message and the dummy message)
and uses it to re-encode his own message. In addition, the noiseless feedback is used to
generate a pair of secret keys $(K_{1}^{*},K_{2}^{*})$, and $K_{j}^{*}$ ($j=1,2$) is used to encrypt the sub-message $W_{j,1}$.

\item Comparing the above code construction of $\mathcal{C}^{in}_{s}$ with that of $\mathcal{C}^{DF}_{s}$,
the encoding and decoding schemes of these two bounds
are almost the same, except that the sub-message $W_{j,1}$ ($j=1,2$) is encrypted by a secret key $K_{j}^{*}$. Thus the secrecy sum rate $R_{1}+R_{2}$
is bounded by two part: the first part is the upper bound on the sum rate of $\mathcal{C}^{DF}_{s}$, and the second part is
the upper bound on the rate of the secret keys $K_{1}^{*}$ and $K_{2}^{*}$.
Using the balanced coloring lemma introduced by
Ahlswede and Cai \cite{AC}, we conclude that the rate of the secret keys $K_{1}^{*}$ and $K_{2}^{*}$ is bounded by
$\min\{H(Y|X_{1},X_{2},Z),I(X_{1},X_{2};Z)\}$.
Thus, the hybrid inner bound $\mathcal{C}^{in}_{s}$ is obtained.

\end{itemize}
The details of the proof
are in Appendix \ref{appen1}.
\end{IEEEproof}

\begin{remark}\label{R1}

Comparing the DF inner bound $\mathcal{C}^{DF}_{s}$ and the partial DF inner bound $\mathcal{C}^{PDF}_{s}$ with our hybrid new inner bound $\mathcal{C}^{in}_{s}$,
it is easy to see that our new inner bound $\mathcal{C}^{in}_{s}$
is strictly larger than
$\mathcal{C}^{DF}_{s}$ and $\mathcal{C}^{PDF}_{s}$.
\end{remark}

\begin{theorem}\label{Tk}
For the discrete memoryless MAC-WT with noiseless feedback,
an outer bound $\mathcal{C}^{out}_{s}$ on the secrecy capacity region $\mathcal{C}_{s}$ is given by
\begin{eqnarray*}
&&\mathcal{C}^{out}_{s}=\{(R_{1}\geq 0,R_{2}\geq 0): R_{1}+R_{2}\leq H(Y|Z)\},
\end{eqnarray*}
for some distribution
\begin{eqnarray}\label{the1}
&&P_{Z,Y|X_{1},X_{2}}(z,y|x_{1},x_{2})\cdot P_{X_{1}X_{2}}(x_{1},x_{2}).
\end{eqnarray}
\end{theorem}

\begin{IEEEproof}
The outer bound $\mathcal{C}^{out}_{s}$ is a simple sato-type outer bound, and the proof is in Appendix \ref{appen2}.
\end{IEEEproof}

\section{Gaussian Example}\label{secIII}
\setcounter{equation}{0}

\subsection{Capacity Results on the Gaussian MAC-WT with Noiseless Feedback}\label{secIII-1}

For the Gaussian case of the model of Figure \ref{f1}, the channel inputs and outputs satisfy
\begin{eqnarray}\label{e301}
&&Y=X_{1}+X_{2}+N_{1}\,\,\,Z=X_{1}+X_{2}+N_{2},
\end{eqnarray}
where 
the channel noises $N_{1}$ and $N_{2}$ are independent and Gaussian distributed, 
i.e., $N_{1}\sim \mathcal{N}(0,\sigma_{1}^{2})$, and $N_{2}\sim \mathcal{N}(0,\sigma_{2}^{2})$.
The average power constraint of the transmitted signal
$X_{j}$ ($j=1,2$) is given by
\begin{eqnarray}\label{e302}
&&\frac{1}{N}\sum_{i=1}^{N}E[X^{2}_{ji}]\leq P_{j}, \,\, j=1,2.
\end{eqnarray}

\emph{\textbf{The DF and partial DF inner bounds on the secrecy capacity region for the Gaussian case of the model of Figure \ref{f1}:}}

\begin{theorem}\label{Ts1}
The DF inner bound $C^{gdf}_{s}$ and the partial DF inner bound $C^{gpdf}_{s}$ for the Gaussian case of the model of Figure \ref{f1} are given by
\begin{eqnarray}\label{e303.1}
&&C^{gdf}_{s}=C^{gpdf}_{s}=\{(R_{1}\geq 0,R_{2}\geq 0): R_{1}\leq \frac{1}{2}\log(1+\frac{P_{1}}{\sigma_{1}^{2}}),\nonumber\\
&&R_{2}\leq \frac{1}{2}\log(1+\frac{P_{2}}{\sigma_{1}^{2}}),\nonumber\\
&&R_{1}+R_{2}\leq \frac{1}{2}\log(1+\frac{P_{1}+P_{2}}{\sigma_{1}^{2}})-\frac{1}{2}\log(1+\frac{P_{1}+P_{2}}{\sigma_{2}^{2}})\}.
\end{eqnarray}
\end{theorem}
\begin{IEEEproof}
In Remark \ref{R1.1}, we have shown that for the model of Figure \ref{f1}, the DF inner bound is the same as the partial DF inner bound.
Along the lines of \cite[pp. 610-611]{tang}, we have
\begin{eqnarray}\label{e303.1.1}
&&C^{gdf}_{s}=C^{gpdf}_{s}=\{(R_{1}\geq 0,R_{2}\geq 0): R_{1}\leq \frac{1}{2}\log(1+\frac{P_{1}}{\sigma_{1}^{2}}),\nonumber\\
&&R_{2}\leq \frac{1}{2}\log(1+\frac{P_{2}}{\sigma_{1}^{2}}),\nonumber\\
&&R_{1}+R_{2}\leq \min\{\frac{1}{2}\log(1+\frac{P_{1}+P_{2}}{\sigma_{1}^{2}}), \frac{1}{2}\log(1+\frac{P_{1}}{\sigma_{1}^{2}})
+\frac{1}{2}\log(1+\frac{P_{2}}{\sigma_{1}^{2}})\}\nonumber\\
&&-\frac{1}{2}\log(1+\frac{P_{1}+P_{2}}{\sigma_{2}^{2}})\}.
\end{eqnarray}
Note that in (\ref{e303.1.1}), $\frac{1}{2}\log(1+\frac{P_{1}+P_{2}}{\sigma_{1}^{2}})\leq \frac{1}{2}\log(1+\frac{P_{1}}{\sigma_{1}^{2}})
+\frac{1}{2}\log(1+\frac{P_{2}}{\sigma_{1}^{2}})$, and thus (\ref{e303.1}) is obtained. The proof is completed.
\end{IEEEproof}

\emph{\textbf{The hybrid inner bound on the secrecy capacity region for the Gaussian case of the model of Figure \ref{f1}:}}
\begin{theorem}\label{Ts2}
The hybrid inner bound $C^{gi}_{s}$ for the Gaussian case of the model of Figure \ref{f1} is given by
\begin{eqnarray}\label{e303.xb}
&&C^{gi}_{s}=\{(R_{1}\geq 0,R_{2}\geq 0): R_{1}\leq \frac{1}{2}\log(1+\frac{P_{1}}{\sigma_{1}^{2}}),\nonumber\\
&&R_{2}\leq \frac{1}{2}\log(1+\frac{P_{2}}{\sigma_{1}^{2}}),\nonumber\\
&&R_{1}+R_{2}\leq \frac{1}{2}\log(1+\frac{P_{1}+P_{2}}{\sigma_{1}^{2}})-\frac{1}{2}\log(1+\frac{P_{1}+P_{2}}{\sigma_{2}^{2}})\nonumber\\
&&+\min\{\frac{1}{2}\log(2\pi e\sigma_{1}^{2}), \frac{1}{2}\log(1+\frac{P_{1}+P_{2}}{\sigma_{2}^{2}})\}\}.
\end{eqnarray}
\end{theorem}
\begin{IEEEproof}
Similar to the corresponding proof in \cite[pp. 610-611]{tang}, substituting $X_{1}=\sqrt{(1-\alpha)P_{1}}U+\sqrt{\alpha P_{1}}U_{1}$ ($0\leq \alpha\leq 1$)
and $X_{2}=\sqrt{(1-\beta)P_{2}}U+\sqrt{\beta P_{2}}U_{2}$ ($0\leq \beta\leq 1$) into (\ref{e301}), and using the fact that
$U$, $U_{1}$ and $U_{2}$ are independent and Gaussian distributed with zero mean and unit variance, and 
$\frac{1}{2}\log(1+\frac{P_{1}+P_{2}}{\sigma_{1}^{2}})\leq \frac{1}{2}\log(1+\frac{P_{1}}{\sigma_{1}^{2}})
+\frac{1}{2}\log(1+\frac{P_{2}}{\sigma_{1}^{2}})$,
(\ref{e303.xb}) is directly obtained. Here note that (\ref{e303.xb}) is achieved when $\alpha=1$ and $\beta=1$. The proof is completed.
\end{IEEEproof}

\emph{\textbf{The outer bound on the secrecy capacity region for the Gaussian case of the model of Figure \ref{f1}:}}
\begin{theorem}\label{Ts3}
For the case that $\sigma_{1}^{2}\geq\sigma_{2}^{2}$, the outer bound $C^{go}_{s}$ for the Gaussian case of the model of Figure \ref{f1} is given by
\begin{eqnarray}\label{e303.xb1}
&&C^{go}_{s}=\{(R_{1}\geq 0,R_{2}\geq 0): R_{1}+R_{2}\leq \frac{1}{2}\log(2\pi e(\sigma_{1}^{2}-\sigma_{2}^{2}))\}.
\end{eqnarray}
For the case that $\sigma_{1}^{2}\leq\sigma_{2}^{2}$, the outer bound $C^{go}_{s}$ is given by
\begin{eqnarray}\label{e303.xb2}
&&C^{go}_{s}=\{(R_{1}\geq 0,R_{2}\geq 0): R_{1}+R_{2}\leq \frac{1}{2}\log(2\pi e(\sigma_{2}^{2}-\sigma_{1}^{2}))
+\frac{1}{2}\log\frac{P_{1}+P_{2}+\sigma_{1}^{2}}{P_{1}+P_{2}+\sigma_{2}^{2}}\}.
\end{eqnarray}
\end{theorem}
\begin{IEEEproof}
\begin{itemize}
\item For the case that $\sigma_{1}^{2}\geq\sigma_{2}^{2}$, (\ref{e301}) can be re-written as
\begin{eqnarray}\label{e301.xxx1}
&&Y=X_{1}+X_{2}+N_{2}+N_{1}-N_{2}\,\,\,Z=X_{1}+X_{2}+N_{2}.
\end{eqnarray}
Substituting (\ref{e301.xxx1}) into Theorem \ref{Tk}, we have
\begin{eqnarray}\label{e303.xb2mm}
&&R_{1}+R_{2}\leq h(Y|Z)=h(X_{1}+X_{2}+N_{2}+N_{1}-N_{2}|X_{1}+X_{2}+N_{2})\nonumber\\
&&=h(N_{1}-N_{2}|X_{1}+X_{2}+N_{2})\leq h(N_{1}-N_{2})=\frac{1}{2}\log(2\pi e(\sigma_{1}^{2}-\sigma_{2}^{2})).
\end{eqnarray}
\item For the case that $\sigma_{1}^{2}\leq\sigma_{2}^{2}$, (\ref{e301}) can be re-written as
\begin{eqnarray}\label{e301.xxx2}
&&Y=X_{1}+X_{2}+N_{1}\,\,\,Z=X_{1}+X_{2}+N_{1}+N_{2}-N_{1}.
\end{eqnarray}
Substituting (\ref{e301.xxx2}) into Theorem \ref{Tk}, we have
\begin{eqnarray}\label{e303.xb2mm1}
&&R_{1}+R_{2}\leq h(Y|Z)=h(Y,Z)-h(Z)=h(Z|Y)+h(Y)-h(Z)\nonumber\\
&&=h(X_{1}+X_{2}+N_{1}+N_{2}-N_{1}|X_{1}+X_{2}+N_{1})+h(Y)-h(Y+N_{2}-N_{1})\nonumber\\
&&=h(N_{2}-N_{1}|X_{1}+X_{2}+N_{1})+h(Y)-h(Y+N_{2}-N_{1})\nonumber\\
&&\leq h(N_{2}-N_{1})+h(Y)-h(Y+N_{2}-N_{1})\nonumber\\
&&\stackrel{(a)}\leq h(N_{2}-N_{1})+h(Y)-\frac{1}{2}\log(2^{2h(Y)}+2^{2h(N_{2}-N_{1})})\nonumber\\
&&\stackrel{(b)}\leq h(N_{2}-N_{1})+\frac{1}{2}\log(2\pi e(P_{1}+P_{2}+\sigma_{1}^{2}))
-\frac{1}{2}\log(2\pi e(P_{1}+P_{2}+\sigma_{1}^{2})+2\pi e(\sigma_{2}^{2}-\sigma_{1}^{2}))\nonumber\\
&&=\frac{1}{2}\log(2\pi e(\sigma_{2}^{2}-\sigma_{1}^{2}))+\frac{1}{2}\log(2\pi e(P_{1}+P_{2}+\sigma_{1}^{2}))
-\frac{1}{2}\log(2\pi e(P_{1}+P_{2}+\sigma_{1}^{2})+2\pi e(\sigma_{2}^{2}-\sigma_{1}^{2}))\nonumber\\
&&=\frac{1}{2}\log(2\pi e(\sigma_{2}^{2}-\sigma_{1}^{2}))+\frac{1}{2}\log\frac{P_{1}+P_{2}+\sigma_{1}^{2}}{P_{1}+P_{2}+\sigma_{2}^{2}},
\end{eqnarray}
where (a) is from the entropy power inequality, i.e., $2^{2h(Y+N_{2}-N_{1})}\geq 2^{2h(Y)}+2^{2h(N_{2}-N_{1})}$,
and (b) is from the fact that $h(Y)-\frac{1}{2}\log(2^{2h(Y)}+2^{2h(N_{2}-N_{1})})$ is increasing while $h(Y)$ is increasing, 
$h(Y)=h(X_{1}+X_{2}+N_{1})\leq \frac{1}{2}\log(2\pi e(P_{1}+P_{2}+\sigma_{1}^{2}))$ and 
$h(N_{2}-N_{1})=\frac{1}{2}\log(2\pi e(\sigma_{2}^{2}-\sigma_{1}^{2}))$.
\end{itemize}
The proof is completed.
\end{IEEEproof}

Finally, recall that Tekin and Yener \cite{TY1} have shown that for the Gaussian MAC-WT without feedback, an inner bound
$C^{gmac-wt}_{s}$ is given by
\begin{eqnarray}\label{e303.2}
&&C^{gmac-wt}_{s}=\{(R_{1}\geq 0,R_{2}\geq 0): R_{1}\leq \frac{1}{2}\log(1+\frac{P_{1}}{\sigma_{1}^{2}})
-\frac{1}{2}\log(1+\frac{P_{1}}{\sigma_{2}^{2}+P_{2}}),\nonumber\\
&&R_{2}\leq \frac{1}{2}\log(1+\frac{P_{2}}{\sigma_{1}^{2}})-\frac{1}{2}\log(1+\frac{P_{2}}{\sigma_{2}^{2}+P_{1}}),\nonumber\\
&&R_{1}+R_{2}\leq \frac{1}{2}\log(1+\frac{P_{1}+P_{2}}{\sigma_{1}^{2}})-\frac{1}{2}\log(1+\frac{P_{1}+P_{2}}{\sigma_{2}^{2}})\}.
\end{eqnarray}
For the case that $\sigma_{1}^{2}\leq\sigma_{2}^{2}$,
the following Figure \ref{f2} shows the inner bound $C^{gi}_{s}$, the partial ($C^{gpdf}_{s}$) and full ($C^{gdf}_{s}$)
DF inner bounds for the Gaussian case of Figure \ref{f1}, the outer bound $C^{go}_{s}$
and Tekin-Yener's inner bound $C^{gmac-wt}_{s}$ of the Gaussian MAC-WT \cite{TY1}
for $P_{1}=P_{2}=1$, $\sigma_{1}^{2}=1$ and $\sigma_{2}^{2}=10$. From Figure \ref{f2}, it is easy to see that our new inner bound $C^{gi}_{s}$ 
is larger than the DF inner bounds $C^{gpdf}_{s}$
and $C^{gdf}_{s}$, and the noiseless feedback helps to enhance the secrecy rate region $C^{gmac-wt}_{s}$ of
the Gaussian MAC-WT.

For the case that $\sigma_{1}^{2}\geq\sigma_{2}^{2}$, the DF bounds $C^{gpdf}_{s}$, $C^{gdf}_{s}$
and Tekin-Yener's inner bound $C^{gmac-wt}_{s}$ reduce to the point $(R_{1}=0, R_{2}=0)$.
The following Figure \ref{f2.x} shows the inner bound $C^{gi}_{s}$ and the outer bound $C^{go}_{s}$
for $P_{1}=P_{2}=10$, $\sigma_{1}^{2}=5$, $\sigma_{2}^{2}=2$. It is easy to see that when $\sigma_{1}^{2}\geq\sigma_{2}^{2}$, 
our hybrid inner bound still provides positive secrecy rates, while there is no positive secrecy rate in the partial and full DF inner bounds.

\begin{figure}[htb]
\centering
\includegraphics[scale=0.5]{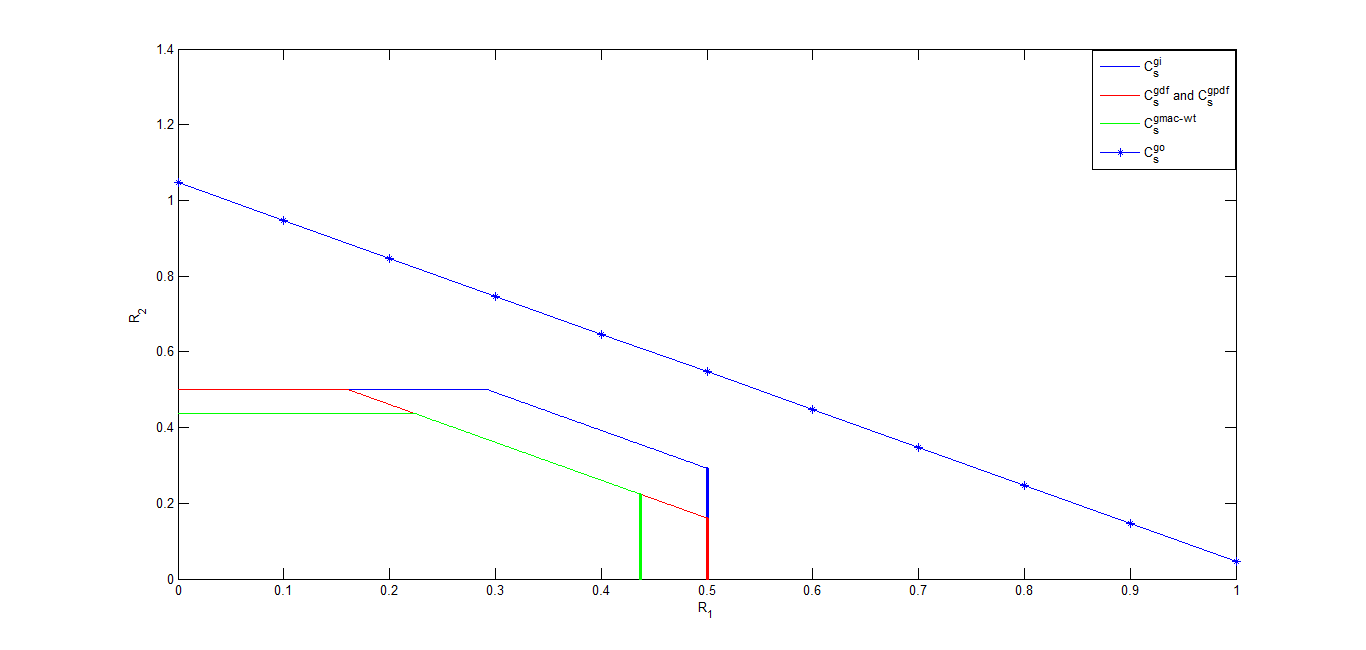}
\caption{The bounds $C^{gi}_{s}$, $C^{gpdf}_{s}$, $C^{gdf}_{s}$, $C^{go}_{s}$,
and $C^{gmac-wt}_{s}$ for $P_{1}=P_{2}=1$, $\sigma_{1}^{2}=1$, $\sigma_{2}^{2}=10$}
\label{f2}
\end{figure}

\begin{figure}[htb]
\centering
\includegraphics[scale=0.5]{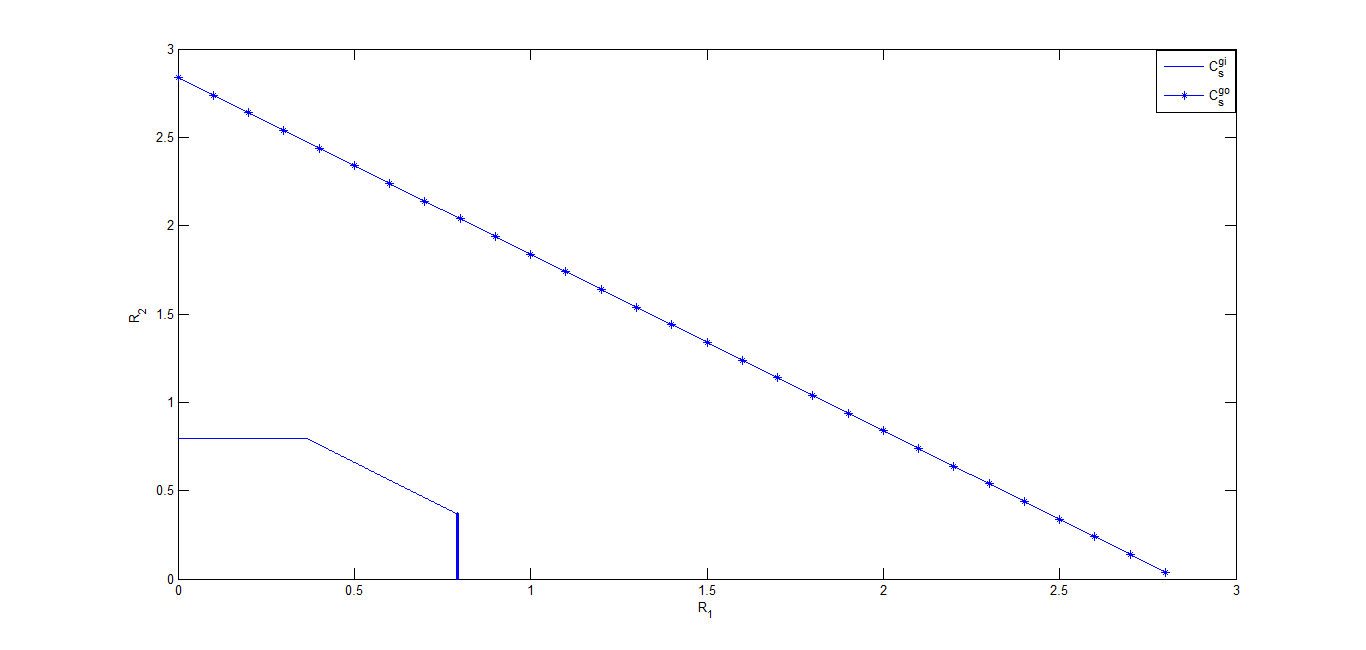}
\caption{The bounds $C^{gi}_{s}$ and $C^{go}_{s}$ for $P_{1}=P_{2}=10$, $\sigma_{1}^{2}=5$, $\sigma_{2}^{2}=2$}
\label{f2.x}
\end{figure}

\subsection{Power Control for the Maximum Secrecy Sum Rate of $\mathcal{C}^{gi}_{s}$}\label{secIII-2}

In this subsection, we assume that the average power constraints of the transmitters
satisfy
\begin{eqnarray}\label{dota1}
&&0\leq P_{1}, P_{2}\leq P,
\end{eqnarray}
and define the maximum secrecy sum rate $R^{*}_{sum}$ of $C^{gi}_{s}$ as
\begin{eqnarray}\label{dota2}
&&R^{*}_{sum}=\max_{P_{1},P_{2}}\frac{1}{2}\log(1+\frac{P_{1}+P_{2}}{\sigma_{1}^{2}})-\frac{1}{2}\log(1+\frac{P_{1}+P_{2}}{\sigma_{2}^{2}})\nonumber\\
&&+\min\{\frac{1}{2}\log(2\pi e\sigma_{1}^{2}), \frac{1}{2}\log(1+\frac{P_{1}+P_{2}}{\sigma_{2}^{2}})\}.
\end{eqnarray}
In the remainder of this subsection, we calculate the maximum secrecy sum rate $R^{*}_{sum}$ of $C^{gi}_{s}$, and show the 
optimum power control (the optimum of $P_{1}$ and $P_{2}$ is denoted by $P_{1}^{*}$ and $P_{2}^{*}$, respectively) for $R^{*}_{sum}$.

\begin{theorem}\label{Ts4}
If $\sigma_{1}^{2}>\sigma_{2}^{2}$, the maximum secrecy sum rate $R^{*}_{sum}$ of $C^{gi}_{s}$ is given by
\begin{equation}\label{dota3}
R^{*}_{sum}=
\left\{
\begin{array}{ll}
\frac{1}{2}\log(1+\frac{2P}{\sigma_{1}^{2}}), & 0\leq P\leq \frac{(2\pi e\sigma_{1}^{2}-1)\sigma_{2}^{2}}{2}\\
\frac{1}{2}\log(1+\frac{(2\pi e\sigma_{1}^{2}-1)\sigma_{2}^{2}}{\sigma_{1}^{2}}), & P\geq \frac{(2\pi e\sigma_{1}^{2}-1)\sigma_{2}^{2}}{2},
\end{array}
\right.
\end{equation}
and the optimum power control is given by
\begin{equation}\label{dota4}
(P_{1}^{*},P_{2}^{*})=
\left\{
\begin{array}{ll}
(P,P), & 0\leq P\leq \frac{(2\pi e\sigma_{1}^{2}-1)\sigma_{2}^{2}}{2}\\
(\frac{(2\pi e\sigma_{1}^{2}-1)\sigma_{2}^{2}}{2},\frac{(2\pi e\sigma_{1}^{2}-1)\sigma_{2}^{2}}{2}), & P\geq \frac{(2\pi e\sigma_{1}^{2}-1)\sigma_{2}^{2}}{2}.
\end{array}
\right.
\end{equation}
If $\sigma_{1}^{2}\leq\sigma_{2}^{2}$, the maximum secrecy sum rate $R^{*}_{sum}$ of $C^{gi}_{s}$ is given by
\begin{equation}\label{dota5}
R^{*}_{sum}=
\left\{
\begin{array}{ll}
\frac{1}{2}\log(1+\frac{2P}{\sigma_{1}^{2}}), & 0\leq P\leq \frac{(2\pi e\sigma_{1}^{2}-1)\sigma_{2}^{2}}{2}\\
\frac{1}{2}\log(2\pi e\sigma_{1}^{2})+\frac{1}{2}\log(1+\frac{2P}{\sigma_{1}^{2}})-\frac{1}{2}\log(1+\frac{2P}{\sigma_{2}^{2}}), & P\geq \frac{(2\pi e\sigma_{1}^{2}-1)\sigma_{2}^{2}}{2},
\end{array}
\right.
\end{equation}
and the optimum power control is given by
\begin{equation}\label{dota6}
(P_{1}^{*},P_{2}^{*})=
\left\{
\begin{array}{ll}
(P,P), & 0\leq P\leq \frac{(2\pi e\sigma_{1}^{2}-1)\sigma_{2}^{2}}{2}\\
(P,P), & P\geq \frac{(2\pi e\sigma_{1}^{2}-1)\sigma_{2}^{2}}{2}.
\end{array}
\right.
\end{equation}
\end{theorem}
\begin{IEEEproof}
From Theorem \ref{Ts2}, it is easy to see that the secrecy sum rate $R_{sum}$ of $C^{gi}_{s}$ is given by
\begin{eqnarray}\label{dota7}
&&R_{sum}=\frac{1}{2}\log(1+\frac{P_{1}+P_{2}}{\sigma_{1}^{2}})-\frac{1}{2}\log(1+\frac{P_{1}+P_{2}}{\sigma_{2}^{2}})
+\min\{\frac{1}{2}\log(2\pi e\sigma_{1}^{2}), \frac{1}{2}\log(1+\frac{P_{1}+P_{2}}{\sigma_{2}^{2}})\},
\end{eqnarray}
and (\ref{dota7}) can be re-written as
\begin{equation}\label{dota8}
R_{sum}=
\left\{
\begin{array}{ll}
\frac{1}{2}\log(1+\frac{P_{1}+P_{2}}{\sigma_{1}^{2}}), & 0\leq P_{1}+P_{2}\leq (2\pi e\sigma_{1}^{2}-1)\sigma_{2}^{2}\\
\frac{1}{2}\log(1+\frac{P_{1}+P_{2}}{\sigma_{1}^{2}})-\frac{1}{2}\log(1+\frac{P_{1}+P_{2}}{\sigma_{2}^{2}})
+\frac{1}{2}\log(2\pi e\sigma_{1}^{2}), & P_{1}+P_{2}>(2\pi e\sigma_{1}^{2}-1)\sigma_{2}^{2}.
\end{array}
\right.
\end{equation}
Since $0\leq P_{1}+P_{2}\leq 2P$, the secrecy sum rate $R_{sum}$ in (\ref{dota8}) can be considered into the following three cases:
\begin{itemize}
\item (Case 1:) If $0\leq P\leq \frac{(2\pi e\sigma_{1}^{2}-1)\sigma_{2}^{2}}{2}$,
it is easy to see that $R_{sum}$ is increasing while $P_{1}$ and $P_{2}$ are increasing, and thus
we have $R^{*}_{sum}=\frac{1}{2}\log(1+\frac{2P}{\sigma_{1}^{2}})$, and the corresponding
optimum $P^{*}_{1}$ and $P^{*}_{2}$ equal to $P$.

\item (Case 2:) If $P>\frac{(2\pi e\sigma_{1}^{2}-1)\sigma_{2}^{2}}{2}$ and $\sigma_{1}^{2}\leq \sigma_{2}^{2}$,
(\ref{dota8}) is re-written as
\begin{equation}\label{dota10}
R_{sum}=
\left\{
\begin{array}{ll}
\frac{1}{2}\log(1+\frac{P_{1}+P_{2}}{\sigma_{1}^{2}}), & 0\leq P_{1}+P_{2}\leq (2\pi e\sigma_{1}^{2}-1)\sigma_{2}^{2}\\
\frac{1}{2}\log(1+\frac{P_{1}+P_{2}}{\sigma_{1}^{2}})-\frac{1}{2}\log(1+\frac{P_{1}+P_{2}}{\sigma_{2}^{2}})
+\frac{1}{2}\log(2\pi e\sigma_{1}^{2}), & (2\pi e\sigma_{1}^{2}-1)\sigma_{2}^{2}<P_{1}+P_{2}\leq 2P.
\end{array}
\right.
\end{equation} 
It is not difficult to show that for this case, $R^{*}_{sum}=\frac{1}{2}\log(1+\frac{2P}{\sigma_{1}^{2}})-\frac{1}{2}\log(1+\frac{2P}{\sigma_{2}^{2}})
+\frac{1}{2}\log(2\pi e\sigma_{1}^{2})$, and the corresponding
optimum $P^{*}_{1}$ and $P^{*}_{2}$ equal to $P$.

\item (Case 3:) If $P>\frac{(2\pi e\sigma_{1}^{2}-1)\sigma_{2}^{2}}{2}$ and $\sigma_{1}^{2}>\sigma_{2}^{2}$, it
is not difficult to show that for this case, $R^{*}_{sum}=\frac{1}{2}\log(1+\frac{(2\pi e\sigma_{1}^{2}-1)\sigma_{2}^{2}}{\sigma_{1}^{2}})$, 
and the corresponding
optimum $P^{*}_{1}$ and $P^{*}_{2}$ equal to $\frac{(2\pi e\sigma_{1}^{2}-1)\sigma_{2}^{2}}{2}$.
\end{itemize}
Combining the above three cases, Theorem \ref{Ts4} is obtained, and the proof is completed.
\end{IEEEproof}

The following Figure \ref{f2.xx} and Figure \ref{f2.xxx} show the maximum secrecy sum rate $R^{*}_{sum}$ 
and the corresponding optimum power control for $\sigma_{1}^{2}>\sigma_{2}^{2}$ and $\sigma_{1}^{2}\leq\sigma_{2}^{2}$, respectively. 
It is easy to see that for both cases, $R^{*}_{sum}$ tends to a constant while $P$ tends to infinity.

\begin{figure}[htb]
\centering
\includegraphics[scale=0.5]{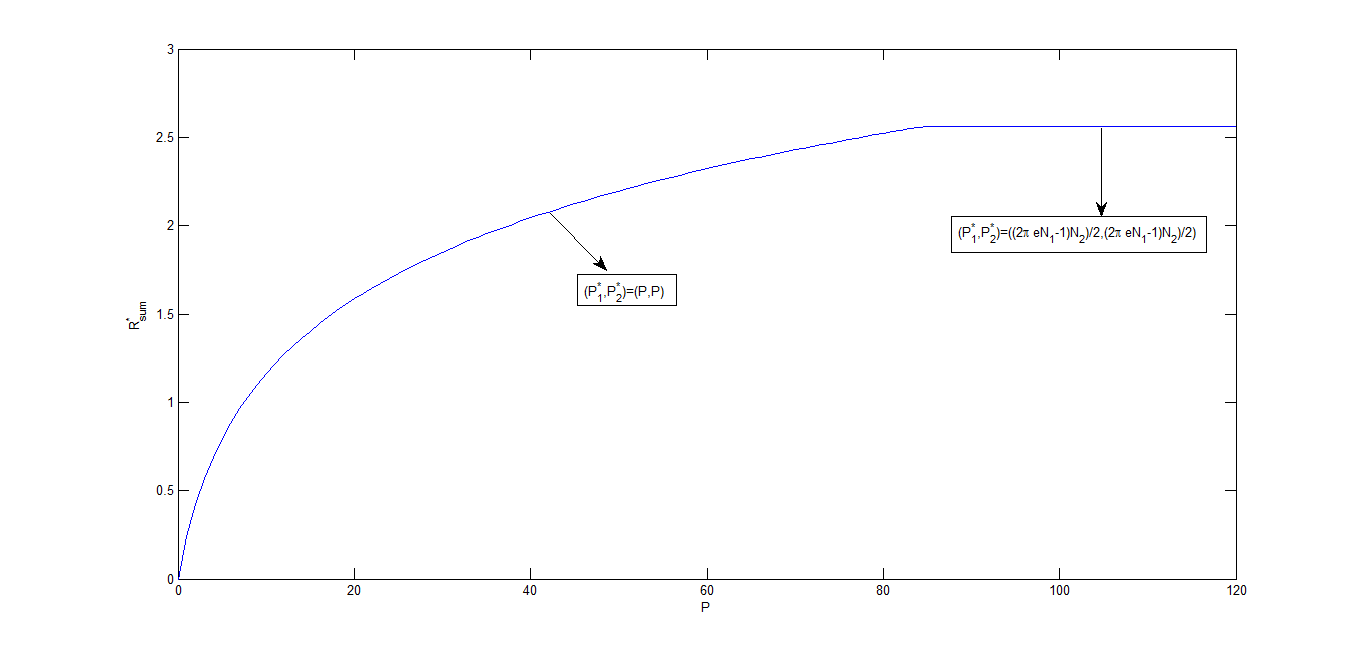}
\caption{The maximum secrecy sum rate $R^{*}_{sum}$ and the corresponding optimum power control for $\sigma_{1}^{2}=5$ and $\sigma_{2}^{2}=2$}
\label{f2.xx}
\end{figure}

\begin{figure}[htb]
\centering
\includegraphics[scale=0.5]{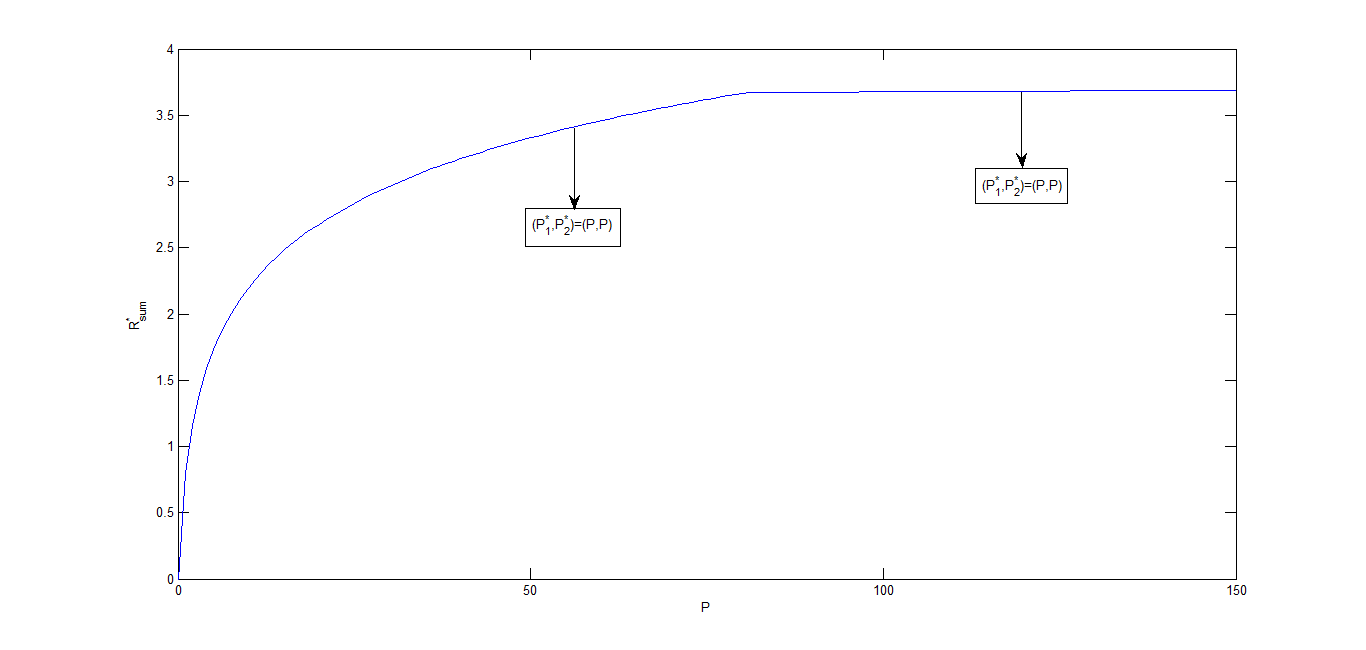}
\caption{The maximum secrecy sum rate $R^{*}_{sum}$ and the corresponding optimum power control for $\sigma_{1}^{2}=1$ and $\sigma_{2}^{2}=10$}
\label{f2.xxx}
\end{figure}

\section{Conclusions\label{secIV}}

In this paper, we present two inner bounds and one outer bound on the secrecy capacity region of the MAC-WT with noiseless feedback. 
To be specific, the first inner bound is constructed
by using the DF strategy, where each transmitter decodes the other one's transmitted message from the noiseless feedback
and then uses the decoded message to re-encode his own messages. The second inner bound is constructed by combining
Ahlswede and Cai's idea of generating
a secret key from the noiseless feedback \cite{AC} with the DF strategy used in the first inner bound.
The outer bound is a simple sato-type bound.
We show that the second inner bound is strictly larger than the first one, and the capacity results are further explained via a Gaussian example.

\section*{Acknowledgement}

This work was supported by
the National Natural Science Foundation of China under Grants 61671391, 61301121 and 61571373,
the fundamental research funds for the Central universities (No.
2682016ZDPY06),
and the Open Research Fund of the State Key Laboratory of Integrated Services Networks, Xidian University (No. ISN17-13).

\renewcommand{\theequation}{A\arabic{equation}}
\appendices\section{Proof of Theorem \ref{T1}\label{appen1}}
\setcounter{equation}{0}

The messages $W_{1}=(W_{1,1},...,W_{1,n})$ and $W_{2}=(W_{2,1},...,W_{2,n})$ are transmitted through $n$ blocks.
In block $i$ ($1\leq i\leq n$),
the transmitted message $w_{j,i}$ ($j=1,2$) is denoted by $w_{j,i}=(w_{j,i,0},w_{j,i,1})$, where
$w_{j,i,0}\in \{1,2,...,2^{NR_{j0}}\}$, $w_{j,i,1}\in \{1,2,...,2^{NR_{j1}}\}$, $w_{j,i}\in \{1,2,...,2^{NR_{j}}\}$ and $R_{j}=R_{j0}+R_{j1}$.
Here note that in block $1$, the transmitted message $w_{j,1}=(w_{j,1,0},const)$ ($j=1,2$), which implies that the sub-message $w_{j,1,1}$ is a constant.
For block $i$ ($1\leq i\leq n$), let $w_{1,i}^{*}$ and $w_{2,i}^{*}$ be the randomly generated dummy messages for transmitters 1 and 2, respectively. Here
$w_{j,i}^{*}\in \{1,2,...,2^{NR^{*}_{j}}\}$ ($j=1,2$).

For $1\leq i\leq n$, let $\widetilde{X}_{j,i}$ ($j=1,2$), $\widetilde{U}_{i}$, $\widetilde{Y}_{i}$ and $\widetilde{Z}_{i}$ be the random vectors with length $N$ for block $i$.
The specific values of the above random vectors are denoted by lower case letters. Moreover, let $X_{j}^{n}=(\widetilde{X}_{j,1},...,\widetilde{X}_{j,n})$,
$U^{n}=(\widetilde{U}_{1},...,\widetilde{U}_{n})$, $Y^{n}=(\widetilde{Y}_{1},...,\widetilde{Y}_{n})$ and $Z^{n}=(\widetilde{Z}_{1},...,\widetilde{Z}_{n})$.

\emph{Construction of the code-books}: In each block $i$ ($1\leq i\leq n$),
for a fixed joint probability $P_{Z,Y|X_{1},X_{2}}(z,y|x_{1},x_{2})\\P_{X_{1}|U}(x_{1}|u)P_{X_{2}|U}(x_{2}|u)P_{U}(u)$,
randomly generate $2^{N(R_{10}+R_{11}+R^{*}_{1}+R_{20}+R_{21}+R^{*}_{2})}$ i.i.d. sequences $\widetilde{u}_{i}$ according to $P_{U}(u)$, and index
these sequences as $\widetilde{u}_{i}(w^{'}_{0,i})$, where
$1\leq w^{'}_{0,i}\leq 2^{N(R_{10}+R_{11}+R^{*}_{1}+R_{20}+R_{21}+R^{*}_{2})}$.

For each $w^{'}_{0,i}$, randomly generate $2^{N(R_{j0}+R_{j1}+R^{*}_{j})}$ ($j=1,2$) i.i.d. sequences
$\widetilde{x}_{j,i}$ according to $P_{X_{j}|U}(x_{j}|u)$, and index these sequences as $\widetilde{x}_{j,i}(w_{j,i}^{'})$, where
$1\leq w_{j,i}^{'}\leq 2^{N(R_{j0}+R_{j1}+R^{*}_{j})}$.

\emph{Encoding scheme}: In block $1$, both the transmitters choose $w^{'}_{0,1}=1$ as the index of the transmitted $\widetilde{u}_{1}$,
and send $\widetilde{u}_{1}(1)$. Furthermore, the transmitter $j$ ($j=1,2$) chooses
$w_{j,1}^{'}=(w_{j,1,0},w_{j,1,1}=const,w_{j,1}^{*})$ as the index of the transmitted codeword $\widetilde{x}_{j,1}$.

In block $i$ ($2\leq i\leq n$), suppose that transmitter 1 has already obtained $w^{'}_{0,i-1}$ and $w_{1,i-1}^{'}=(w_{1,i-1,0},w_{1,i-1,1},w_{1,i-1}^{*})$.
Since the transmitter 1 receives the feedback $\widetilde{y}_{i-1}$, he tries to find a unique
sequence $\widetilde{x}_{2,i-1}(\check{w}_{2,i-1}^{'},w^{'}_{0,i-1})$ such that $(\widetilde{x}_{2,i-1}(\check{w}_{2,i-1}^{'},w^{'}_{0,i-1}),
\widetilde{x}_{1,i-1}(w_{1,i-1}^{'},w^{'}_{0,i-1}),\widetilde{u}_{i-1}(w^{'}_{0,i-1}),\\ \widetilde{y}_{i-1})$ are jointly typical sequences.
From AEP, it is easy to see that the error probability $Pr\{\check{w}_{2,i-1}^{'}\neq w_{2,i-1}^{'}\}$ goes to $0$ if
\begin{eqnarray}\label{app1}
&&R_{20}+R_{21}+R^{*}_{2}\leq I(X_{2};Y|X_{1},U).
\end{eqnarray}
Thus in block $i$, the transmitter $1$ sends $\widetilde{u}_{i}$ with the index $w^{'}_{0,i}=(w_{1,i-1}^{'},\check{w}_{2,i-1}^{'})$.

Analogously, since the transmitter 2 receives the feedback $\widetilde{y}_{i-1}$, he tries to find a unique
sequence $\widetilde{x}_{1,i-1}(\tilde{w}_{1,i-1}^{'},w^{'}_{0,i-1})$ such that $(\widetilde{x}_{1,i-1}(\tilde{w}_{1,i-1}^{'},w^{'}_{0,i-1}),
\widetilde{x}_{2,i-1}(w_{2,i-1}^{'},w^{'}_{0,i-1}),\widetilde{u}_{i-1}(w^{'}_{0,i-1}),\\ \widetilde{y}_{i-1})$ are jointly typical sequences.
From AEP, it is easy to see that the error probability $Pr\{\tilde{w}_{1,i-1}^{'}\neq w_{1,i-1}^{'}\}$ goes to $0$ if
\begin{eqnarray}\label{app2}
&&R_{10}+R_{11}+R^{*}_{1}\leq I(X_{1};Y|X_{2},U).
\end{eqnarray}
Thus in block $i$, the transmitter $2$ sends $\widetilde{u}_{i}$ with the index $w^{'}_{0,i}=(\tilde{w}_{1,i-1}^{'},w_{2,i-1}^{'})$.

In block $i$ ($2\leq i\leq n$), before choosing the transmitted codewords $\widetilde{x}_{1,i}$ and $\widetilde{x}_{2,i}$,
we generate a mapping
$g_{i}: \widetilde{y}_{i-1}\rightarrow \{1,2,...,2^{N(R_{11}+R_{21})}\}$. Furthermore, we define $K_{i}^{*}=(K_{i,1}^{*},K_{i,2}^{*})=g_{i}(\widetilde{Y}_{i-1})$
as a random variable uniformly distributed over $\{1,2,...,2^{N(R_{11}+R_{21})}\}$, and it is
independent of $\widetilde{X}_{1,i}$, $\widetilde{X}_{2,i}$,
$\widetilde{Y}_{i}$, $\widetilde{Z}_{i}$, $W_{1,i}$, $W_{2,i}$, $W_{1,i}^{*}$ and $W_{2,i}^{*}$. Here note that $K_{i,j}^{*}$ ($j=1,2$) is used as
a secret key shared by the transmitter $j$ and the receiver, and $k_{i,j}^{*}\in \{1,2,...,2^{NR_{j1}}\}$ is a specific value of $K_{i,j}^{*}$.
Reveal the mapping $g_{i}$ to the transmitters, receiver and the eavesdropper.
After the generation of the secret key, the transmitter $j$ ($j=1,2$) sends $\widetilde{x}_{j,i}$ with the index
$w_{j,i}^{'}=(w_{j,i,0},w_{j,i,1}\oplus k_{i,j}^{*},w_{j,i}^{*})$.

\emph{Decoding scheme for the receiver}: The intended receiver does backward decoding after the transmission of all $n$ blocks is completed,
and the receiver's decoding scheme is exactly the same as that of the classical MAC with feedback \cite[pp. 295-296]{coverz}.
Following similar steps of error probability analysis for MAC with feedback \cite[pp. 295-296]{coverz}, we have
\begin{eqnarray}\label{app3}
&&R_{10}+R_{11}+R^{*}_{1}+R_{20}+R_{21}+R^{*}_{2}\leq I(X_{1},X_{2};Y).
\end{eqnarray}
\emph{Equivocation analysis (1): For block $2\leq i\leq n$, a lower bound on $H(K_{i}^{*}|\widetilde{X}_{1,i-1},\widetilde{X}_{2,i-1},\widetilde{Z}_{i-1})$}:
Given $\widetilde{X}_{1,i-1}$, $\widetilde{X}_{2,i-1}$ and $\widetilde{Z}_{i-1}$, the eavesdropper's equivocation
about the secret key $k_{i}^{*}$ can be bounded by Ahlswede and Cai's balanced coloring lemma \cite[p. 260]{AC}, see the followings.
\begin{lemma}\label{Lx}
\textbf{(Balanced coloring lemma)} For
arbitrary $\epsilon, \delta>0$, sufficiently large $N$, all $N$-type
$P_{X_{1}X_{2}Y}(x_{1},x_{2},y)$ and all
$\widetilde{x}_{1,i-1}, \widetilde{x}_{2,i-1}\in T_{X_{1}X_{2}}^{N}$ ($2\leq i\leq n$),
there exists a $\gamma$- coloring $c: T_{Y|X_{1},X_{2}}^{N}(\widetilde{x}_{1,i-1}, \widetilde{x}_{2,i-1})\rightarrow \{1,2,..,\gamma\}$
of $T_{Y|X_{1},X_{2}}^{N}(\widetilde{x}_{1,i-1}, \widetilde{x}_{2,i-1})$
such that for all joint $N$-type $P_{X_{1}X_{2}YZ}(x_{1},x_{2},y,z)$ with marginal distribution \\$P_{X_{1}X_{2}Z}(x_{1},x_{2},z)$ and
$\frac{|T_{Y|X_{1},X_{2},Z}^{N}(\widetilde{x}_{1,i-1},\widetilde{x}_{2,i-1},\widetilde{z}_{i-1})|}{\gamma}\geq 2^{N\epsilon}$,
$\widetilde{x}_{1,i-1}, \widetilde{x}_{2,i-1}, \widetilde{z}_{i-1}\in T_{X_{1}X_{2}Z}$,
\begin{equation}\label{app4}
|c^{-1}(k)|\leq \frac{|T_{Y|X_{1},X_{2},Z}^{N}(\widetilde{x}_{1,i-1}, \widetilde{x}_{2,i-1},\widetilde{z}_{i-1})|(1+\delta)}{\gamma},
\end{equation}
for $k=1,2,...,\gamma$, where $c^{-1}$ is the inverse image of $c$.
\end{lemma}
From Lemma \ref{Lx}, we see that the typical set $T_{Y|X_{1},X_{2},Z}^{N}(\widetilde{x}_{1,i-1},\widetilde{x}_{2,i-1},\widetilde{z}_{i-1})$ maps into at least
\begin{eqnarray}\label{app5}
&&\frac{|T_{Y|X_{1},X_{2},Z}^{N}(\widetilde{x}_{1,i-1},\widetilde{x}_{2,i-1},\widetilde{z}_{i-1}))|}
{\frac{|T_{Y|X_{1},X_{2},Z}^{N}(\widetilde{x}_{1,i-1},\widetilde{x}_{2,i-1},\widetilde{z}_{i-1})|(1+\delta)}{\gamma}}=\frac{\gamma}{1+\delta}
\end{eqnarray}
colors. On the other hand, the typical set $T_{Y|X_{1},X_{2},Z}^{N}(\widetilde{x}_{1,i-1},\widetilde{x}_{2,i-1},\widetilde{z}_{i-1})$
maps into at most $\gamma$ colors.
From (\ref{app5}), we can conclude that
\begin{eqnarray}\label{app6}
&&H(K_{i}^{*}|\widetilde{X}_{1,i-1},\widetilde{X}_{2,i-1},\widetilde{Z}_{i-1})\geq \log\frac{\gamma}{1+\delta}.
\end{eqnarray}
Here note that $\frac{|T_{Y|X_{1},X_{2},Z}^{N}(\widetilde{x}_{1,i-1},\widetilde{x}_{2,i-1},\widetilde{z}_{i-1})|}{\gamma}\geq 2^{N\epsilon}$ implies that
$\gamma\leq |T_{Y|X_{1},X_{2},Z}^{N}(\widetilde{x}_{1,i-1},\widetilde{x}_{2,i-1},\widetilde{z}_{i-1})|$.
Choosing $\gamma=|T_{Y|X_{1},X_{2},Z}^{N}(\widetilde{x}_{1,i-1},\widetilde{x}_{2,i-1},\widetilde{z}_{i-1})|$ and noticing that
\begin{eqnarray}\label{app7}
&&|T_{Y|X_{1},X_{2},Z}^{N}(\widetilde{x}_{1,i-1},\widetilde{x}_{2,i-1},\widetilde{z}_{i-1})|
\geq (1-\epsilon_{1})2^{N(1-\epsilon_{2})H(Y|X_{1},X_{2},Z)},
\end{eqnarray}
where $\epsilon_{1}$ and $\epsilon_{2}$ tend to $0$ as $N$ tends to infinity, (\ref{app6}) can be further bounded by
\begin{eqnarray}\label{app8}
&&H(K_{i}^{*}|\widetilde{X}_{1,i-1},\widetilde{X}_{2,i-1},\widetilde{Z}_{i-1})
\geq \log\frac{1-\epsilon_{1}}{1+\delta}+N(1-\epsilon_{1})H(Y|X_{1},X_{2},Z).
\end{eqnarray}

\emph{Equivocation analysis (2): Bound on eavesdropper's equivocation $\Delta$}:
For all blocks, the equivocation $\Delta$ is bounded by
\begin{eqnarray}\label{app9}
&&\Delta=\frac{1}{nN}H(W_{1},W_{2}|Z^{n})\stackrel{(a)}=\frac{1}{nN}(H(W^{'}_{1,0},W^{'}_{2,0}|Z^{n})\nonumber\\
&&+H(W^{'}_{1,1},W^{'}_{2,1}|Z^{n},W^{'}_{1,0},W^{'}_{2,0})),
\end{eqnarray}
where (a) is from the definitions $W^{'}_{j,0}=(W_{j,1,0},...,W_{j,n,0})$ and $W^{'}_{j,1}=(W_{j,2,1},...,W_{j,n,1})$ for $j=1,2$.
The conditional entropy $H(W^{'}_{1,0},W^{'}_{2,0}|Z^{n})$ of (\ref{app9}) is bounded by
\begin{eqnarray}\label{app10}
&&H(W^{'}_{1,0},W^{'}_{2,0}|Z^{n})=H(W^{'}_{1,0},W^{'}_{2,0},Z^{n})-H(Z^{n})\nonumber\\
&&=H(W^{'}_{1,0},W^{'}_{2,0},Z^{n},X_{1}^{n},X_{2}^{n})
-H(X_{1}^{n},X_{2}^{n}|W^{'}_{1,0},W^{'}_{2,0},Z^{n})-H(Z^{n})\nonumber\\
&&\stackrel{(b)}=H(Z^{n}|X_{1}^{n},X_{2}^{n})+H(X_{1}^{n},X_{2}^{n})
-H(X_{1}^{n},X_{2}^{n}|W^{'}_{1,0},W^{'}_{2,0},Z^{n})-H(Z^{n})\nonumber\\
&&\stackrel{(c)}=nN(R_{10}+R_{11}+R^{*}_{1}+R_{20}+R_{21}+R^{*}_{2})
-nNI(X_{1},X_{2};Z)-H(X_{1}^{n},X_{2}^{n}|W^{'}_{1,0},W^{'}_{2,0},Z^{n})\nonumber\\
&&\stackrel{(d)}\geq nN(R_{10}+R_{11}+R^{*}_{1}+R_{20}+R_{21}+R^{*}_{2})
-nNI(X_{1},X_{2};Z)-nN\epsilon_{3},
\end{eqnarray}
where (b) is from $H(W^{'}_{1,0}|X_{1}^{n})=0$ and $H(W^{'}_{2,0}|X_{2}^{n})=0$, (c) is from the code constructions of $X_{1}^{n}$, $X_{2}^{n}$ and
the fact that the channel is memoryless,
and (d) is from the fact that given $w^{'}_{1,0}$, $w^{'}_{2,0}$ and $z^{n}$, the eavesdropper tries to find unique $w^{'}_{1,1}$, $w^{'}_{2,1}$,
$w_{1}^{*}=(w_{1,1}^{*},...,w_{1,n}^{*})$ and $w_{2}^{*}=(w_{2,1}^{*},...,w_{2,n}^{*})$ such that $(x_{1}^{n},x_{2}^{n},z^{n})$ are jointly typical, and
from the properties of AEP, we see that
the eavesdropper's decoding error probability tends to $0$ if
\begin{eqnarray}\label{app11}
&&R_{1,1}+R_{2,1}+R_{1}^{*}+R_{2}^{*}\leq I(X_{1},X_{2};Z),
\end{eqnarray}
then by using Fano's inequality, we have $\frac{1}{nN}H(X_{1}^{n},X_{2}^{n}|W^{'}_{1,0},W^{'}_{2,0},Z^{n})\leq \epsilon_{3}$, where $\epsilon_{3}\rightarrow 0$
as $n, N\rightarrow \infty$.
Moreover, the conditional entropy $H(W^{'}_{1,1},W^{'}_{2,1}|Z^{n},W^{'}_{1,0},W^{'}_{2,0})$ of (\ref{app9}) is bounded by
\begin{eqnarray}\label{app12}
&&H(W^{'}_{1,1},W^{'}_{2,1}|Z^{n},W^{'}_{1,0},W^{'}_{2,0})\nonumber\\
&&\geq \sum_{i=2}^{n}H(W_{1,i,1},W_{2,i,1}|Z^{n},W^{'}_{1,0},W^{'}_{2,0},W_{1,1,1},W_{2,1,1},\nonumber\\
&&...,W_{1,i-1,1},W_{2,i-1,1},W_{1,i,1}\oplus K_{i,1}^{*},W_{2,i,1}\oplus K_{i,2}^{*})\nonumber\\
&&\stackrel{(e)}=\sum_{i=2}^{n}H(W_{1,i,1},W_{2,i,1}|\widetilde{Z}_{i-1},W_{1,i,1}\oplus K_{i,1}^{*},
W_{2,i,1}\oplus K_{i,2}^{*})\nonumber\\
&&\geq \sum_{i=2}^{n}H(W_{1,i,1},W_{2,i,1}|\widetilde{Z}_{i-1},\widetilde{X}_{1,i-1},\widetilde{X}_{2,i-1},
W_{1,i,1}\oplus K_{i,1}^{*},W_{2,i,1}\oplus K_{i,2}^{*})\nonumber\\
&&=\sum_{i=2}^{n}H(K_{i,1}^{*},K_{i,2}^{*}|\widetilde{Z}_{i-1},\widetilde{X}_{1,i-1},\widetilde{X}_{2,i-1},W_{1,i,1}\oplus K_{i,1}^{*},
W_{2,i,1}\oplus K_{i,2}^{*})\nonumber\\
&&\stackrel{(f)}=\sum_{i=2}^{n}H(K_{i}^{*}|\widetilde{Z}_{i-1},\widetilde{X}_{1,i-1},\widetilde{X}_{2,i-1})\nonumber\\
&&\stackrel{(g)}\geq (n-1)(\log\frac{1-\epsilon_{1}}{1+\delta}+N(1-\epsilon_{1})H(Y|X_{1},X_{2},Z)),\nonumber\\
\end{eqnarray}
where (e) is from the Markov chain $(W_{1,i,1},W_{2,i,1})\rightarrow (\widetilde{Z}_{i-1},W_{1,i,1}\oplus K_{i,1}^{*},W_{2,i,1}\oplus K_{i,2}^{*})\\
\rightarrow (W^{'}_{1,0},W^{'}_{2,0},W_{1,1,1},W_{2,1,1},...,W_{1,i-1,1},W_{2,i-1,1},\widetilde{Z}_{1},...,\widetilde{Z}_{i-2},\\
\widetilde{Z}_{i},...,\widetilde{Z}_{n})$, (f) is from the definition $K_{i}^{*}=(K_{i,1}^{*},K_{i,2}^{*})$ and the Markov chain
$K_{i}^{*}\rightarrow (\widetilde{Z}_{i-1},\widetilde{X}_{1,i-1},\widetilde{X}_{2,i-1})\rightarrow (W_{1,i,1}\oplus K_{i,1}^{*}, W_{2,i,1}\oplus K_{i,2}^{*})$,
and (g) is from (\ref{app8}).

Substituting (\ref{app10}) and (\ref{app12}) into (\ref{app9}), we have
\begin{eqnarray}\label{app13}
&&\Delta\geq R_{10}+R_{11}+R^{*}_{1}+R_{20}+R_{21}+R^{*}_{2}
-I(X_{1},X_{2};Z)-\epsilon_{3}
+\frac{n-1}{nN}\log\frac{1-\epsilon_{1}}{1+\delta}\nonumber\\
&&+\frac{n-1}{n}(1-\epsilon_{1})H(Y|X_{1},X_{2},Z).
\end{eqnarray}
The bound (\ref{app13}) implies that if
\begin{eqnarray}\label{app14}
&&R^{*}_{1}+R^{*}_{2}\geq I(X_{1},X_{2};Z)-H(Y|X_{1},X_{2},Z)
\end{eqnarray}
we can prove that $\Delta\geq R_{10}+R_{11}+R_{20}+R_{21}-\epsilon$
by choosing sufficiently large $n$ and $N$.

Finally, applying Fourier-Motzkin elimination (see, e.g., \cite{lall}) on
(\ref{app1}), (\ref{app2}), (\ref{app3}), (\ref{app11}) and (\ref{app14}), Theorem \ref{T1} is obtained.
The proof of Theorem \ref{T1} is completed.

\section{Proof of Theorem \ref{Tk}\label{appen2}}

Note that
\begin{eqnarray}\label{b1}
R_{1}+R_{2}-\epsilon&\stackrel{(1)}\leq& \frac{H(W_{1},W_{2}|Z^{N})}{N}\nonumber\\
&=&\frac{1}{N}(H(W_{1},W_{2}|Z^{N})-H(W_{1},W_{2}|Z^{N},Y^{N})+H(W_{1},W_{2}|Z^{N},Y^{N}))\nonumber\\
&\stackrel{(2)}\leq&\frac{1}{N}(I(W_{1},W_{2};Y^{N}|Z^{N})+\delta(P_{e}))\nonumber\\
&\leq&\frac{1}{N}(H(Y^{N}|Z^{N})+\delta(P_{e}))\nonumber\\
&=&\frac{1}{N}\sum_{i=1}^{N}H(Y_{i}|Y^{i-1},Z^{N})+\frac{\delta(P_{e})}{N}\nonumber\\
&\leq&\frac{1}{N}\sum_{i=1}^{N}H(Y_{i}|Z_{i})+\frac{\delta(P_{e})}{N}\nonumber\\
&\stackrel{(3)}=&\frac{1}{N}\sum_{i=1}^{N}H(Y_{i}|Z_{i},J=i)+\frac{\delta(P_{e})}{N}\nonumber\\
&\stackrel{(4)}=&H(Y_{J}|Z_{J},J)+\frac{\delta(P_{e})}{N}\nonumber\\
&\stackrel{(5)}\leq&H(Y_{J}|Z_{J})+\frac{\delta(\epsilon)}{N}\nonumber\\
&\stackrel{(6)}=&H(Y|Z)+\frac{\delta(\epsilon)}{N},
\end{eqnarray}
where (1) is from (\ref{e205}), (2) is from Fano's inequality, 
(3) and (4) are from the fact that $J$ is a random variable (uniformly distributed
over $\{1,2,...,N\}$), and it is independent of $Y^{N}$, $Z^{N}$, $W_{1}$ and $W_{2}$,
(5) is from $P_{e}\leq \epsilon$ and
$\delta(P_{e})$ is increasing while $P_{e}$ is increasing, and (6) is from the definitions $Y\triangleq Y_{J}$ and $Z\triangleq Z_{J}$.
Letting $\epsilon\rightarrow 0$,
$R_{1}+R_{2}\leq H(Y|Z)$ is proved.
The proof of Theorem \ref{Tk} is completed.


\begin{thebibliography}{99}

\bibitem{Wy} A. D. Wyner, ``The wire-tap channel,"
{\sl The Bell System Technical Journal}, vol. 54, no. 8, pp.
1355-1387, 1975.

\bibitem{CK} I. Csisz$\acute{a}$r and J. K\"{o}rner, ``Broadcast channels with confidential messages," {\sl IEEE Trans.
Inf. Theory}, vol. IT-24, no. 3, pp. 339-348, May 1978.


\bibitem{supp1} J. M. Wozencraft, M. Horstein, ``Coding for two-way channels,'' {\sl MASSACHUSETTS INST OF
TECH CAMBRIDGE RESEARCH LAB OF ELECTRONICS}, 1961.

\bibitem{coverx} T. M. Cover and J. A. Thomas,
{\sl Elements of Information Theory}. New York, NY: Wiley-Interscience, 1991.

\bibitem{coverz} T. M. Cover and C. S. K. Leung,
``An achievable rate region for the multiple-access channel with feedback,"
{\sl IEEE Trans. Inf. Theory}, vol. IT-27, no. 3, pp. 292-298, 1981.

\bibitem{CG1} T. M. Cover and A. El Gamal, ``Capacity theorems for the relay channel,''
{\sl IEEE Trans. Inf. Theory}, vol. IT-25, pp. 572-584, 1979.


\bibitem{lapidoth} I. S. Bross and A. Lapidoth, ``An improved achievable region for the discrete memoryless
two-user multiple-access channel with noiseless feedback," {\sl IEEE Trans.
Inf. Theory}, vol. IT-51, no. 3, pp. 811-833, 2005.


\bibitem{AC} R. Ahlswede and N. Cai, ``Transmission, Identification and Common Randomness Capacities for
Wire-Tap Channels with Secure Feedback from the Decoder," book
chapter in {\sl General Theory of Information Transfer and
Combinatorics}, LNCS 4123,  pp. 258-275, Berlin: Springer-Verlag,
2006.

\bibitem{AFJK} E. Ardestanizadeh, M. Franceschetti, T.Javidi and Y.Kim, ``Wiretap channel with secure rate-limited feedback,"
{\sl IEEE Trans. Inf. Theory}, vol. IT-55, no. 12, pp. 5353-5361,
December 2009.

\bibitem{dai1} B. Dai, A. J. Han vinck, Y. Luo and Z. Zhuang, ``Capacity region of non-degraded wiretap
channel with noiseless feedback,'' {\sl Proceedings of 2012 IEEE International Symposium on Information Theory}, USA, 2012.

\bibitem{dai2} B. Dai, Z. Ma and X. Fang, ``Feedback Enhances the Security of State-Dependent
Degraded Broadcast Channels With Confidential Messages,''
{\sl IEEE Trans. Inf. Forensics and Security}, Vol. 10, No. 7, pp. 1529-1542, 2015.

\bibitem{TY1} E. Tekin and A. Yener, ``The Gaussian multiple access wire-tap channel,"
{\sl IEEE Trans. Inf. Theory}, vol. IT-54, no. 12, pp. 5747-5755, Dec. 2008.

\bibitem{TY2} E. Tekin and A. Yener, ``The general Gaussian multiple access and
two-way wire-tap channels: Achievable rates and cooperative jamming,"
{\sl IEEE Trans. Inf. Theory}, vol. IT-54, no. 6, pp. 2735-2751, June 2008.


\bibitem{EU} E. Ekrem and S. Ulukus, ``On the secrecy of multiple access wiretap
channel," in {\sl Proc. Annual Allerton Conf. on Communications, Control
and Computing}, Monticello, IL, Sept. 2008.

\bibitem{aref} M. H. Yassaee and M. R. Aref, ``Multiple access wiretap channels with strong secrecy,''
{\sl Proceedings of 2010 IEEE Information Theory Workshop}, pp. 1-5, 2010.


\bibitem{WB} M. Wiese and H. Boche, ``An Achievable Region for the Wiretap Multiple-Access Channel with Common Message,''
{\sl Proceedings of 2012 IEEE International Symposium on Information Theory}, 2012.

\bibitem{tang} X. Tang, R. Liu, P. Spasojevi$\acute{c}$ and H. V. Poor, ``Multiple access channels with generalized feedback and confidential messages,''
{\sl Proceedings of 2007 IEEE Information Theory Workshop}, pp. 608-613, 2007.


\bibitem{lall} S. Lall, ``Advanced topics in computation for control,'' Lecture notes for
Engr210b, Stanford University, Fall, 2004.








\end{thebibliography}
\end{document}